# ESO observing programme: VST Early-type GAlaxy Survey (VEGAS)

## Abstract

VEGAS is a deep multi-band ($u'g'r'i'$) imaging survey, carried out with the ESO VLT Survey Telescope (VST). VST is a 2.6-m wide-field optical survey telescope, located at ESO Paranal Observatory (Chile). First VEGAS observations started in October 2011 (former PI: M. Capaccioli, see also Capaccioli et al. 2015). Later the program was approved for an extension to the period 2016-2021 (PI: E. Iodice). The whole VEGAS sample is made by selecting groups and clusters of galaxies with an early-type galaxy in the core brighter than $M_B$ = −21 mag, in the local volume within 54 Mpc/h, mainly located in the Southern hemisphere. The total observing time allocated to the survey is 500 hours for five years (2016-2021). With the data release 1 (DR1), we provide the reduced VST mosaics of 10 targets, which have been presented in the VEGAS publications.

Taking advantage of the wide (1 deg$^2$) field-of-view of OmegaCAM@VST, the long integration time and the wide variety of targets, VEGAS has proven to be a gold mine to explore the structure of galaxies down to the faintest surface brightness levels of ∼ 27-30 mag/arcsec$^2$ in the SDSS $g'$-band, for the dense clusters of galaxies as well as for the unexplored poor groups of galaxies. As such, in the wide panorama of deep imaging surveys, VEGAS has occupied a pivotal role in exploring the galaxy properties as a function of the environments down to the low surface brightness (LSB) regime. About 30% of the VEGAS observing time was dedicated to the Fornax Deep Survey (FDS), a new multi- band deep imaging survey of the Fornax cluster, where the reduced data have been recently released (Peletier et al. 2020, arXiv:2008.12633).

To date, using about 400 hours of the total observing time, VEGAS has already collected 43 targets (groups and clusters of galaxies) covering a total area on the sky of ∼ 95 deg$^2$. Based on the analyzed data, VEGAS allowed us to *i)* study the galaxy outskirts, detect the intra-cluster light and LSB features in the intra-cluster/group space (Iodice et al. 2016, 2017a; Spavone et al. 2018; Cattapan et al. 2019; Raj et al. 2019, 2020; Iodice et al. 2019a, 2020a), *ii)* trace the mass assembly in galaxies, by estimating the accreted mass fraction in the stellar halos and provide results that can be directly compared with the predictions of galaxy formation models (Iodice et al. 2017b; Spavone et al. 2017, 2020), *iii)* trace the spatial distribution of candidate globular clusters (D'Abrusco et al. 2016; Cantiello et al. 2018, 2020); *iv)* detect the ultra-diffuse galaxies (Forbes et al. 2019, 2020; Iodice et al. 2020b).

## Overview of Observations

Targets, covered area, filters and total exposure times of the DR1 are listed in **Table 1**. In this table, the adopted observing strategy for each target is also included. We have tested that for the brightest and most extended galaxies (with $m_B$ ≤ 10 mag and a major axis diameter ≥ 3 arcmin), the best background estimate is achieved by adopting the step-dither observing strategy. This mimics the ON-OFF procedure devised in infrared astronomy where the background is estimated from exposures taken as close as possible, in space and time, to the scientific ones. Therefore, the step-dither strategy used for the VEGAS images consists of a cycle of short exposures (150 sec) on the science target and on an adjacent field (close in space and time) to the science frame. We adopted an offset of ≤ 0.3 deg in the observing sequence and the directions of these small offsets were randomly chosen around the center of each field. An average sky image, for each night, is derived from the sky frames, which is then scaled and subtracted from the science frames.
For less extended objects (with a major axis diameter D ≤ 3 arcmin), we adopted the standard diagonal observing strategy, since the sky background can be estimated on the science frame, by using a polynomial surface fit over the entire frame (see Capaccioli et al. 2015).

# Release Content

The DR1 of VEGAS consists of 23 science files and 23 weight maps. The total data volume is ~30 GB. The target list is provided in **Table 1**.

| Target (1) | RA [h m s] (2) | Dec. [d m s] (3) | u' [sec] (4) | g' [sec] (5) | r' [sec] (6) | i' [sec] (7) | Area [deg²] (8) | Strategy (9) |
|---|---|---|---|---|---|---|---|---|
| **IC 1459** | 22:57:10.6 | -36:27:44 | --- | 7125 | 7200 | --- | 2.9 | step-dither |
| **NGC 1533** | 04:09:51.8 | -56:07:06 | --- | 7800 | 4800 | --- | 1.4 | standard |
| **NGC 3115** | 10:05:14.0 | +07:43:07 | 14800 | 8675 | --- | 6030 | 1.2 | standard |
| **NGC 3379** | 10:47:49.6 | +12:34:54 | 7350 | 7650 | 7425 | --- | 3.9 | step-dither |
| **NGC 3923** | 11:51:01.7 | -28:48:22 | --- | 3862 | --- | 3787 | 6.5 | step-dither |
| **NGC 4365** | 12:24:28.3 | +07:19:04 | --- | 6378 | --- | 3620 | 1.4 | standard |
| **NGC 4472** | 12:29:46.7 | +08:00:02 | --- | 5695 | --- | 4423 | 1.6 | standard |
| **NGC 5018** | 13:13:01.0 | -19:31:05 | 7350 | 6225 | 6000 | --- | 3.9 | step-dither |
| **NGC 5044** | 13:15:24.0 | -16:23:08 | --- | 6112 | --- | 1537 | 5.8 | step-dither |
| **NGC 5846** | 15:06:29.3 | +01:36:20 | --- | 2812 | --- | 412 | 5.3 | step-dither |

**Table 1:** Target list of the VEGAS DR1. In column 1 is given the target name. In columns 2 and 3 are listed the J2000 celestial coordinates. From columns 4 to 7 are reported the total integration time for each 1 square deg field, in the *u'*, *g'*, *r'*, and *i'* bands respectively. In column 8 is indicated the total covered area of the mosaic. In column 9 is indicated the adopted observing strategy.

# Release Notes

## Data Reduction and Calibration

Raw VEGAS data in the DR1 were processed with the VST-*tube* pipeline (Grado et al. 2012). A detailed description of all data-reduction steps is given by Capaccioli et al. (2015). In short, they include:

1. pre-reduction
2. astrometric and photometric calibration
3. mosaic production

In the pre-reduction process, science images are treated to remove the instrumental signatures, applying overscan, bias, and flat-field corrections, as well as gain harmonization of the 32 CCDs, illumination correction and, for the *i'*-band, defringing. The absolute photometric calibration is performed by comparing the OmegaCAM magnitudes of the standard star fields observed during each night with SDSS DR8 photometry. For each night and band, the zero point (ZP) and color term were obtained using the tool Photcal (Radovich et al. 2004). The extinction coefficient was derived from the extinction curve M.OMEGACAM.2011-12-01T16:15:04.474 provided by ESO. Relative photometric correction among the exposures was obtained by minimizing the quadratic sum of magnitude differences between overlapping detections, by using the SCAMP task (Bertin 2006). The final coadded images were then normalized to an exposure time of one second and a ZP = 30 mag.

To obtain the absolute and relative astrometric calibrations we used the SCAMP task. For the absolute astrometric calibration, we refer to the 2MASS catalog. Finally, the image resampling, where the astrometric solution is applied, and the final image coaddition are made with SWARP (Bertin et al. 2002).

As an additional task, the VST-tube pipeline can provide sky-subtracted mosaics. For images obtained with the standard observing technique (using diagonal dithers), the sky background is modelled fitting a surface, typically a 2D polynomial, to the pixel values of the mosaic, where all bright sources are masked. The mask is made by using the ExAM task (Huang et al. 2011), a program based on SExtractor (Bertin & Arnouts 1996), which was developed to accurately mask background and foreground sources, as well as reflection haloes and spikes from saturated stars.

For the images acquired with the *step-dither* observing strategy, the background is estimated from exposures taken as sky frames. For each observing night, the pipeline produces an average sky frame which is scaled and subtracted to each science frame.

### Data Quality

In **Table 2** we report the limiting magnitudes and the average FWHM within the field, for each set of observations and in the different photometric bands. Same information is also reported in the image header. The limiting magnitude is the surface brightness of a point source corresponding at 5σ of the background noise in the image. The RMS error of the astrometric solution is ~0.3 arcsec.

| Target | FWHM [arcsec] | | | | depth [mag] | | | |
|---|---|---|---|---|---|---|---|---|
| | u' | g' | r' | i' | u' | g' | r' | i' |
| (1) | (2) | (3) | (4) | (5) | (6) | (7) | (8) | (9) |
| **IC 1459** | --- | 1.73 | 0.89 | --- | --- | 25.4±0.6 | 24.9±0.7 | --- |
| **NGC 1533** | --- | 0.79 | 0.78 | --- | --- | 25.6±0.4 | 24.7±0.3 | --- |
| **NGC 3115** | 0.82 | 1.00 | --- | 0.90 | 25.0±1.0 | 25.4±0.6 | --- | 23.5±0.9 |
| **NGC 3379** | 0.77 | 1.00 | 0.81 | --- | 24.0±0.9 | 25.0±1.0 | 24.5±1.0 | --- |
| **NGC 3923** | --- | 1.14 | --- | 0.99 | --- | 25.4±0.6 | --- | 23.5±0.9 |
| **NGC 4365** | --- | 0.86 | --- | 0.79 | --- | 25.6±0.4 | --- | 23.0±0.7 |
| **NGC 4472** | --- | 0.86 | --- | 0.73 | --- | 25.4±0.6 | --- | 23.5±0.9 |
| **NGC 5018** | 0.77 | 0.77 | 0.94 | --- | 24.0±0.8 | 25.0±1.0 | 24.5±1.0 | --- |
| **NGC 5044** | --- | 1.37 | --- | 1.02 | --- | 25.4±0.5 | --- | 23.0±0.7 |
| **NGC 5846** | --- | 1.18 | --- | 1.37 | --- | 25.4±0.6 | --- | 22.1±0.4 |

**Table 2:** Data quality of the VEGAS DR1. In column 1 is given the target name. From columns 2 to 5 are reported the average FWHM seeing, in the *u'*, *g'*, *r'*, and *i'* bands, respectively. From columns 6 to 9 are reported the limiting magnitude for a point-source computed at 5σ of the background level, in the *u'*, *g'*, *r'*, and *i'* bands, respectively.

### Known issues

None

# Data Format

## Files Types

The files are in FITS format, with the relevant information in the header. Each science frame is accompanied by a weight frame. All files have been compressed using NASA's HEASARC's fpack routine (https://heasarc.gsfc.nasa.gov/fitsio/fpack/). Files are named based on the target, the filter, the date and time of the observation following the format:

- science images:
  <TARGET>-<FILTER>-<DAY>-<MONTH>-<YEAR>-<YY>h<ZZ>m<KK>s_all.fits.fz

- weight maps:
  <TARGET>-<FILTER>-<DAY>-<MONTH>-<YEAR>-<YY>h<ZZ>m<KK>s_all.weight.fits.fz

# Acknowledgements

The VEGAS data were produced by the VEGAS collaboration, with the support of the VST-data center in Naples, at INAF-Astronomical Observatory of Capodimonte and VST funds. E. Iodice wish to thank ESO for the support received during the upload of the VEGAS data into the ESO Phase 3 Science Archive.

The **VEGAS science team** is made by E. Iodice, P.I. (1), M. Spavone, co- PI (1), M. Capaccioli, former PI, (1) and M. Arnaboldi (2), E. Bannikova (3), S. Brough (4), M. Cantiello (5), A. P. Cooper (6), E.M. Corsini (7), G. D'Ago (8), E. Dalla Bont´a (7), D. De Cicco (8), E. Emsellem (2), J. Falcon-Barroso (FDS co-I, 9), D. Forbes (10), A. Grado (1), L. Greggio (11), E. Held (11), M. Hilker (2), D. Krajnovic (12), A. La Marca (13), S. Mieske (14), N.R. Napolitano (15), M. Paolillo (13), A. Pasquali (16), R. Peletier (FDS co-I, 17), A. Pizzella (7), I. Prandoni (18), R. Ragusa (1), M.A. Raj (19), M. Rejkuba (2), R. Rampazzo (11), P. Schipani (1), C. Spiniello (20), G. van de Ven (FDS co-I, 21).

- (1) INAF - Astronomical Observatory of Capodimonte, Salita Moiariello 16, I80131, Naples, Italy
- (2) European Southern Observatory, Karl-Schwarzschild-Strasse 2, D-85748 Garching bei Muenchen, Germany
- (3) Institute of Radio Astronomy of National Academy of Sciences of Ukraine, Mystetstv 4, UA-61022 Kharkiv, Ukraine
- (4) School of Physics, University of New South Wales, NSW 2052, Australia
- (5) INAF-astronomical Abruzzo Observatory, Via Maggini, 64100, Teramo, Italy
- (6) Institute of Astronomy and Department of Physics, National Tsing Hua University, 101 Kuang-Fu Rd. Sec. 2, Hsinchu 30013, Taiwan
- (7) Dipartimento di Fisica e Astronomia 'G. Galilei', Università di Padova, vicolo dell'Osservatorio 3, I-35122 Padova, Italy
- (8) Instituto de Astrofísica, Facultad de Física, Pontificia Universidad Catolica de Chile, Av. Vicuña Mackenna 4860, 7820436 Macul, Santiago, Chile
- (9) Instituto de Astrofísica de Canarias, Calle Vía Láctea s/n, E-38200 La Laguna, Spain
- (10) Centre for Astrophysics and Supercomputing, Swinburne University of Technology, Hawthorn, Victoria 3122, Australia
- (11) INAF – Astronomical Observatory of Padova, Via dell'Osservatorio 8, I-36012, Asiago (VI), Italy
- (12) Leibniz-Institut für Astrophysik Potsdam (AIP), An der Sternwarte 16, 14482, Potsdam, Germany
- (13) University of Naples "Federico II", via Cinthia 21, Naples 80126, Italy
- (14) European Southern Observatory, Alonso de Cordova 3107, Vitacura, Santiago, Chile
- (15) School of Physics and Astronomy, Sun Yat-sen University Zhuhai Campus, 2 Daxue Road, Tangjia, Zhuhai, Guangdong 519082, China
- (16) Astronomisches Rechen-Institut, Zentrum fur Astronomieder Universitat, Monchhofstr. 12-14, D-69120 Heidelberg, Germany
- (17) Kapteyn Institute, University of Groningen, Landleven 12, 9747 AD Groningen, the Netherlands
- (18) INAF – Istituto di Radioastronomia, via Gobetti 101, 40129 Bologna, Italy

(19) INAF – Astronomical Observatory of Rome, via Frascati 33, 00040 - Monte Porzio Catone, Rome, Italy
(20) Sub-Department of Astrophysics, Department of Physics, University of Oxford, Denys Wilkinson Building, Keble Road, Oxford OX1 3RH, UK
(21) Department of Astrophysics, University of Vienna, Turkenschanzstrasse 17, 1180 Wien, Austria

Data products of DR1 are created from observations collected at the European Organisation for Astronomical Research in the Southern Hemisphere under the following ESO programme(s): 089.B-0607(A), 090.B-0414(B, D), 091.B-0614(A), 092.B-0623(B, C, D), 094.B-0496(D), 095.B-0779(A), 096.B-0582(B), 097.B-0806(A, B), 098.B-0208(A), 099.B-0560(A), and 0100.B-0168(A).

Any publication making use of this data, whether obtained from the ESO archive or via third parties, must include the following acknowledgment:

- *"Based on data products created from observations collected at the European Organisation for Astronomical Research in the Southern Hemisphere under ESO programme(s): 089.B-0607(A), 090.B-0414(B, D), 091.B-0614(A), 092.B-0623(B, C, D), 094.B-0496(D), 095.B-0779(A), 096.B-0582(B), 097.B-0806(A, B), 098.B-0208(A), 099.B-0560(A), and 0100.B-0168(A)."*

If the access to the ESO Science Archive Facility services was helpful for you research, please include the following acknowledgment:

- *"This research has made use of the services of the ESO Science Archive Facility."*

Science data products from the ESO archive may be distributed by third parties, and disseminated via other services, according to the terms of the [Creative Commons Attribution 4.0 International license](). Credit to the ESO origin of the data must be acknowledged, and the file headers preserved.

The published VEGAS papers based on the DR1 are listed below:

- Forbes, D.A.; Dullo, B.T.; Gannon, J. et al. 2020, MNRAS, 494, 5293: *Ultra-diffuse galaxies in the IC 1459 group from the VEGAS survey*
- Iodice E., Spavone M., Cattapan A., et al. 2020, A&A, 635, 3: *VEGAS: VST Early-type GAlaxy Survey. V. IC 1459 group: Mass assembly history in low density environments*
- Forbes D.A., Gannon J., Couch W.J. et al. 2019, A&A, 626, 66: *An ultra diffuse galaxy in the NGC 5846 group from the VEGAS survey*
- Cattapan A., Spavone M., Iodice E. et al. 2019, ApJ, 874, 130: *VEGAS: A VST Early-type GAlaxy Survey. IV. NGC 1533, IC 2038 and IC 2039: an interacting triplet in the Dorado group*
- Spavone M., Iodice E., Capaccioli M., et al. 2018, ApJ, 864, 149: *VEGAS: A VST Early-type GAlaxy Survey. III. Mapping the galaxy structure, interactions and intragroup light in the NGC 5018 group*
- Cantiello M., D'Abrusco R., Spavone M. et al., 2018, A&A, 611, 93: *VEGAS-SSS II: Comparing the globular cluster systems in NGC3115 and NGC1399 using VEGAS and FDS survey data*
- Spavone M., Capaccioli M., Napolitano N.R. et al., 2017, A&A, 603, 38: *VEGAS: A VST Early-type GAlaxy Survey. II. Photometric study of giant ellipticals and their stellar halos*
- Spavone M., Capaccioli M., Napolitano N.R. et al., 2017, The Messenger, vol.170, pg.34: *Unveiling the Nature of Giant Ellipticals and their Stellar Halos with the VST*
- Capaccioli M., Spavone M., Grado A. et al., 2015, A&A, 581, 10: *VEGAS: A VST Early-type GAlaxy Survey. I. Presentation, wide-field surface photometry and substructures in NGC 4472*
- Cantiello M., Capaccioli M., Napolitano N.R. et al., 2015, A&A, 576, 14: *VEGAS-SSS. A VST early-type galaxy survey: analysis of small stellar systems. Testing the methodology on the globular cluster system in NGC 3115*

- Cantiello M., Capaccioli M., Napolitano N.R. et al., 2015, The Messenger, vol.159, pg.46: *VEGAS-SSS: A VST Programme to Study the Satellite Stellar Systems around Bright Early-type Galaxies*

References cited in text:

- Grado et al. 2012, Mem. SAIT Supplement, v.19, p.362
- Iodice et al. 2016, ApJ, 820, 42
- D'Abrusco et al. 2016, ApJL, 819L, 31
- Iodice et al. 2017a, ApJ, 851, 75
- Iodice et al. 2017b, ApJ, 839, 21
- Raj et al. 2019, A&A, 628, 4
- Iodice et al. 2019, A&A, 623, A1
- Raj et al. 2020, A&A, 640, 137
- Iodice et al. 2020b, A&A, 642, 48
- Cantiello et al. 2020, A&A, 639, 136
- Spavone et al. 2020, A&A, 639, 14